# An electronic model for self-assembled hybrid organic/perovskite semiconductors: reverse band edge electronic states ordering and spin-orbit coupling

J. Even[1a)], L. Pedesseau[1], M.-A. Dupertuis[2], J.-M. Jancu[1], and C. Katan[1,3]

[1]Université Européenne de Bretagne, INSA, FOTON, UMR 6082, 35708 Rennes, France
[2]Ecole Polytechnique Fédérale de Lausanne (EPFL), Laboratory of Quantum Optoelectronics, CH-1015 Lausanne, Switzerland
[3]CNRS, Institut des Sciences Chimiques de Rennes, UMR 6226, 35042 Rennes, France

[a)] Electronic mail : jacky.even@insa-rennes.fr

# ABSTRACT

Based on density functional theory, the electronic and optical properties of hybrid organic/perovskite crystals are thoroughly investigated. We consider the mono-crystalline 4F-PEPI as material model and demonstrate the optical process is governed by three active Bloch states at the $\Gamma$ point of the reduced Brillouin zone with a reverse ordering compared to tetrahedrally bonded semiconductors. Giant spin-orbit coupling effects and optical activities are subsequently inferred from symmetry analysis.

# I. INTRODUCTION

Semiconductor optoelectronic devices are based on properties of direct band-gap crystals in the zinc-blende[1-2] and wurtzite[3] phases related to p-like valence-band sates and s-like conduction-band states. Nowadays, there is a growing interest in the development of hybrid inorganic-organic architectures that could enable superior optical functions and greatly enhanced device performances. Among them, self-assembled hybrid organic layered perovskites (SAHOP) structures are emerging as a powerful class of two-dimensional (2D) materials due to their broad technological potentialities for nanophotonics and nanoelectronics[4-9] through perfectly controlled growth and organization. Both chemical insight and optical properties of various SAHOP are well documented. In particular, it has been shown that the optical spectra of lead halide organic semiconductors can be easily tailored as function of the organic group, thus improving luminescence efficiencies and/or tuning the emission wavelength[10]. These systems exhibit extremely large exciton binding energies[11] as well as sharp resonances for the biexciton[12] and triexciton[13] transitions. Excitonic switching and Peierls transitions have also been reported[14]. Furthermore, SAHOP based on lead halides have recently demonstrated enhanced non-linear optical properties in microcavities[7-8], ascribed to strong electro-optical couplings. However, manufacturing SAHOP microcavities is a challenging task because divergent technologies have to be accommodated[8]. Besides, the underlying mechanism behind optical process is poorly understood. Indeed, existing modeling of the optical properties and carrier injection is still scanty and does not capture the subtle interplay between the organic sheets and the semiconductor layer. Available theoretical studies either give a general description of the electronic band structure, limited to the inorganic 2D lattices, mostly using density functional theory (DFT)[15] or focus on the excitonic coupling using effective parameters for the carrier dispersion and dielectric confinement[16]. This work aims at bridging over this lack by means of a thorough description of the relevant electronic states and corresponding optical selection rules of a SAHOP material model based on 4F-PEPI[8,17] ([pFC$_6$H$_5$C$_2$H$_4$NH$_3$]$_2$PbI$_4$).

## II. MODEL

Our study is performed by using the DFT implementation available in the ABINIT package[18], with the PBE gradient correction for exchange-correlation[19] and relativistic, norm-conserving, separable, dual-space Gaussian-type pseudopotentials of Goedecker, Teter, and Hutter for all atoms[20]. The electronic wave-functions are expanded onto a plane-wave basis set with an energy cut-off of 950 eV and a 1x4x4 Monkhorst-Pack grid is used for reciprocal space integration. The spin-orbit coupling (SOC) interaction is taken into account. Although numerous SAHOP systems with lead-halides have been studied, only a few related crystallographic structures are known precisely. In fact, growth of monocrystals for X-ray diffraction is difficult because of the lattice disorder and strain induced by the organic molecule that plays a fundamental role in the dielectric confinement. In a few cases, the cation packing is compatible with the inorganic framework, made of corner-shared PbX$_6$ octaedra (X=Cl, Br, I), leading to a perfectly ordered 2D system built on a single crystal hybrid organic perovskite (SCHOP). This is the case of 4F-PEPI with a monoclinic unit cell[17] as shown in Fig. 1. Experimental lattice parameters (transformed to P21/c: a=16.7 Å, b=8.6 Å, c=8.8 Å, β=110°) and atomic positions[19] were used for the DFT calculations with C-H and N-H bond length fixed to 1.089 and 1.008 Å respectively. As X-ray diffraction pattern of spin-coated films of 4F-PEPI lead to the same interlayer spacing (16.7 Å)[17], the SCHOP 4F-PEPI is believed to offer a good reference framework to investigate the electronic and optical properties of its SAHOP counterpart.

## III. RESULTS AND DISCUSSIONS

Figure 2 shows the band structure of mono-crystalline 4F-PEPI with and without the spin-orbit interaction. No energy dispersion occurs along the Γ-X direction (characterizing the stacking axis in real space), which is an inherent consequence of the dielectric mismatch between $PbI_4$ and the organic sheet. As a result, the density of states close to the band gap exhibits a reduced 2D dimensionality in connection with the observed 2D character of excitons in the SAHOP counterpart[11].

The DFT electronic structure also reveals a direct band-gap character in agreement with the observed luminescence at room temperature[8,17]. The fundamental transitions with and without SOC are of 1.2 and 2.0 eV respectively, to be compared to the measured value of 2.35eV. The band gap is known to be underestimated in DFT ground state computation. This deficiency can be corrected by including many-body effects (GW self-energy correction for the band gap and Bethe Salpeter equation resolution for the exciton) but such calculations are beyond available computational resources for large systems. Despite this shortcoming, the overall conclusions related to the energy band dispersions and symmetries are reliable and can help to build accurately semi-empirical Hamiltonians (e.g. in k.p theory and/or tight-binding approximation) where detailed information of Bloch states and selection rules are required. Similar conclusions can be drawn from the electronic band structure (Figs. 3) of the monoclinic crystal of $[C_5H_{11}NH_3]_2PbI_4$. It is a representative member of a large SCHOP and SAHOP family with alcane chains in the organic layer[14,16,21,22].

In first approximation, the optical absorption near the band-edge can be modeled without SOC (Fig. 2a) by three active Bloch states at the Γ-point: a non-degenerate level for the valence-band maximum (VBM) and two nearly doubly-degenerate levels for the conduction-band minimum (CBM). The associated wave-functions are represented for 4F-PEPI in the bc*-plane in Figs. 4.

The VBM wavefunction is real and confined into the PbI$_4$ lattice and consists of anti-bonding hybridizations of Pb *6s* and I *5p* orbitals. It is associated to the nonpolar irreducible representation B$_g$ of the point group C$_{2h}$. B$_g$ is related to the twofold helical axis parallel to PbI$_4$ planes and transforms under E$_{1/2g}$ into the double group[23]. This twofold axis is a direct outcome of the particular stacking of the organic molecules and do not provide for the VBM states the expected A$_g$ character. The in-plane energy dispersion of VBM is found large and almost isotropic in the entire Brillouin zone (Fig. 2a and 3a), which further evidences the s-character of the wave-functions related to the inorganic semiconductor. The $|S\rangle$ symbol is proposed for SCHOP for the first time in this work (Fig. 4), to represent the Bloch State of the VBM, by analogy to CBM in conventional semiconductors[1-3]. A significant mixing with the molecular electronic states is observed only for the lower lying electronic states. In SAHOP structures, a crystal-field splitting $\Delta_{cr}$ should appear owing to the structural anisotropy between parallel to and normal to the stacking axis. It is associated to the dielectric confinement but also to the non-bonding of the iodine p orbitals along the a axis Neglecting spin-orbit coupling, $\Delta_{cr}$ in 4F-PEPI will split the threefold-degenerate p-derived states into a non-degenerate level and a nearly doubly-degenerate one. It is clearly evidenced in the conduction band structure (Fig. 2a and 3a) where the two first excited Bloch states, namely CBM1 and CBM2, are real and almost degenerated at the Γ-point with an energy splitting of about 35meV for 4F-PEPI. CBM1 and CBM2 correspond respectively to the irreducible polar representations B$_u$ and A$_u$ for 4F-PEPI of group C$_{2h}$ resulting in almost in-plane perpendicular "p"-like states. As seen in Figs. 4, their wave-functions are mainly distributed in the surrounding of the Pb shell. The $|Y\rangle$ and $|Z\rangle$ symbols are proposed for SCHOP in this work (Fig. 4), to represent the Bloch States of the CBM, by analogy to VBM in conventional semiconductors[1-3]. We calculate a crystal field of 1eV that corresponds to the energy spacing between (CBM1+CBM2)/2 and CBM3, the latter state displays a non-bonding character of iodine *p* orbitals. Contrarily to the valence band maximum, the effect of the spin-orbit interaction is huge at the conduction band minimum leading to a large SO splitting $\Delta_{SO}$ between the two first conduction states. This is a direct consequence of the *s* and *p* symmetries in the ground and excited wave-functions. Moreover, the CBM1 and CBM2 wave-functions are mixed by the SOC because $B_u$ and $A_u$ transform into the same irreducible representation $E_{1/2u}$ of the double group[23]. The DFT calculation gives $\Delta_{SO}$=1.2eV and this concords with the experimental value of 0.966 eV of an equivalent SAHOP[24].

For optics, it is noteworthy that non-bonding electrons occur in our calculations at very high energies above the CBM and consequently are not involved in the excitation process. The DFT fundamental transition displays a nearly-perfect transverse electric (TE) character in agreement with experimental results on SAHOP[16]. It is similar to conventional zinc-blende quantum wells[1-3] with $D_{2d}$ point symmetry or würtzite bulk compounds with $C_{6v}$ point symmetry. Symmetry and ordering of the Bloch states are however different for SCHOP. In 4F-PEPI, the organic layer stacking may yield small deviations from in-plane spectral activity. In order to get insight into the optical process, we have calculated without SOC the dipolar matrix elements between the first valence and conduction band states as defined by: $M_{VBM,CBMJ} = \left| \left\langle \psi_{VBM} \left| -i\hbar \frac{\partial}{\partial x_i} \right| \psi_{CBMJ} \right\rangle \right|$, where $x_i$ represents the crystal axis and j the two first excited levels. Fig. 5a shows the polar plot of $M_{VBM,CB1}$ and $M_{VB,CB2}$ in the bc-plane. As expected from symmetry analysis, optical strengths of $M^2_{VBM,CBM1}$ (taken as reference) and $M^2_{VBM,CBM2}$ are maximum for light polarized parallel and perpendicular to the b*-axis respectively. Fig. 5b shows that $M_{VB,CB2}$ is maximum along the c axis perpendicular to the perovskite layers (a* direction). In analogy to tetrahedrally-bonded semiconductors, one can define the Kane matrix element[1-3]: $Pj = \frac{\hbar M_{VBM,CBMJ}}{m_e}$ and corresponding energy: $E_{pj} = \frac{2m_e P^2}{\hbar^2}$.

For 4F-PEPI, the mean value of $E_{p_1}$ is of about 5.5eV and this explains the robustness of optoelectronic properties in SAHOP materials. Indeed, $E_p$ amounts typically to 20eV$^2$ for III-V semiconductors which are however characterized by extended Wannier-Mott excitons where binding energies are of few meV. Conversely, it has been demonstrated in SAHOP systems, excitons strongly localize in the inorganic layers constituted of PbX$_6^{4-}$ octahedra due to the dielectric confinement[12,16], thus leading to very large binding energies of few hundreds of meV. Both the sizeable $E_p$ and exciton binding energy cooperatively contribute to generate appealing optical responses in SAHOP. Moreover, the optical anisotropy is small as shown by the ratio of Kane energies in 4F-PEPI: $E_{P2}/E_{P1} = 95\%$. As a consequence, the TE character of fundamental transition in a 4F-PEPI crystal is enforced. It has been clearly demonstrated[16] in another SCHOP ( $(C_{10}H_{21}NH_3)_2PbI_4$ ) materials but it is expected to be a general property of the SCHOP family. A similar behavior is also anticipated for SAHOP counterparts and analogues, the TE character being further enhanced by the orientational disorder introduced by the organic layer, as observed in smectic liquid crystals.

Finally, a schematic representation of the SAHOP electronic band diagram without ($\Delta_{SO}=0$, left) and with ($\Delta_{SO}\neq 0$, right) SOC is proposed in figure 6. This diagram is analogous to the one of III-V semiconductors in würtzite[3] but with a reverse band edge electronic states ordering. The $|Y\rangle$ (CBM1) and $|Z\rangle$ (CBM2) real wave-functions without SOC (left part of figure 5) are mixed by the SOC (right part of figure 6) into real and imaginary parts of the complex spinorial CBM components. The resulting complex spinorial components (for example: $|Y\uparrow\rangle + |iZ\downarrow\rangle$ in figure 6) can be predicted on the basis of a symmetry analysis of the Hamiltonian with and without the SOC effect, but this development analogous to the ones in conventional semiconductors is beyond the scope of the present work.

## IV. CONCLUSION

In summary, based on DFT calculations and symmetry analysis of the Bloch states, the ordering of band edge states and the optical activity including Kane's energy of a prototype SCHOP crystal have been investigated. The significant value of the Kane's energy associated with the large excitonic binding energy reported for the corresponding SAHOP[11] account for the attractive optical responses evidenced in this class of hybrid materials. Additional insight can be gained from a k.p model starting from symmetry properties of the Bloch states which can efficiently model excitonic and spin-orbit couplings.

## ACKNOWLEDGMENTS

This work was performed using HPC resources from GENCI-CINES/IDRIS grant 2012-c2012096724. The work is supported by Agence Nationale pour la Recherche (PEROCAI project ANR-10-04). We gratefully acknowledge Prof. M. Rikukawa and Dr. Y. Takeoka for providing unpublished atomic positions of 4F-PEPI[17].

**FIGURE 1:** (J. Even for PRB)

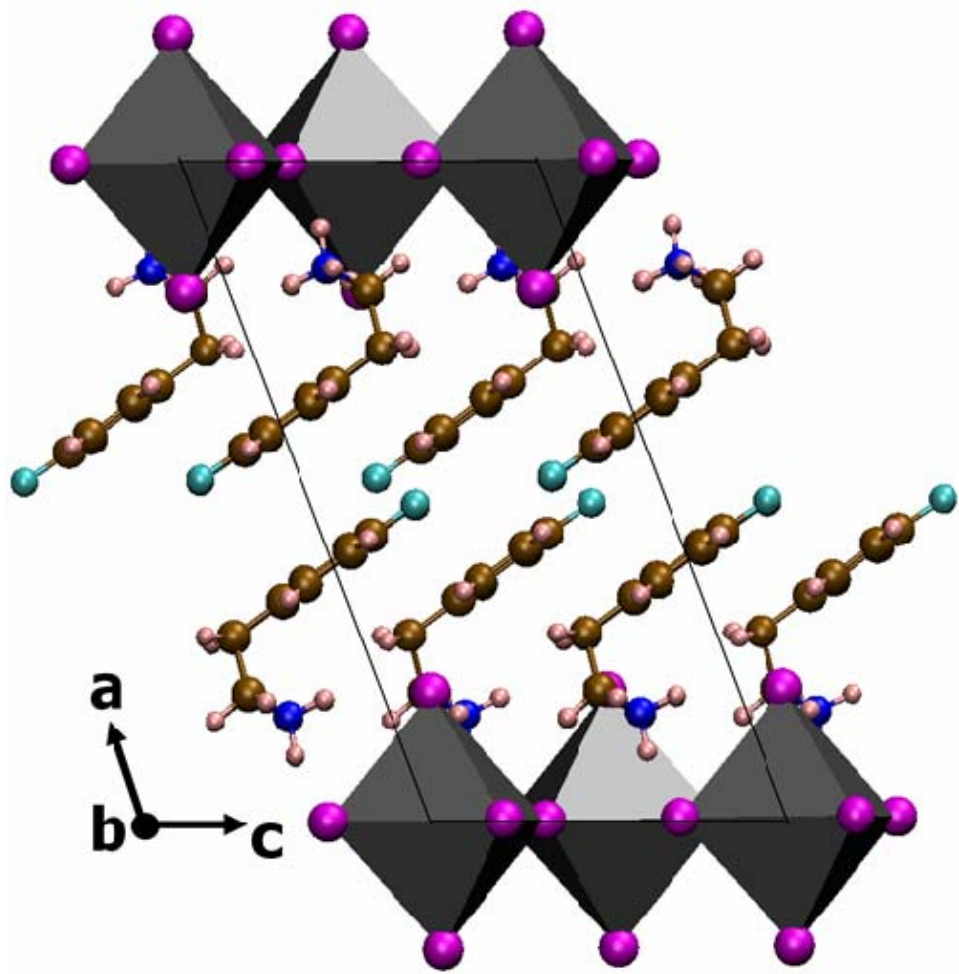

**FIGURE 2:** (J. Even for PRB)

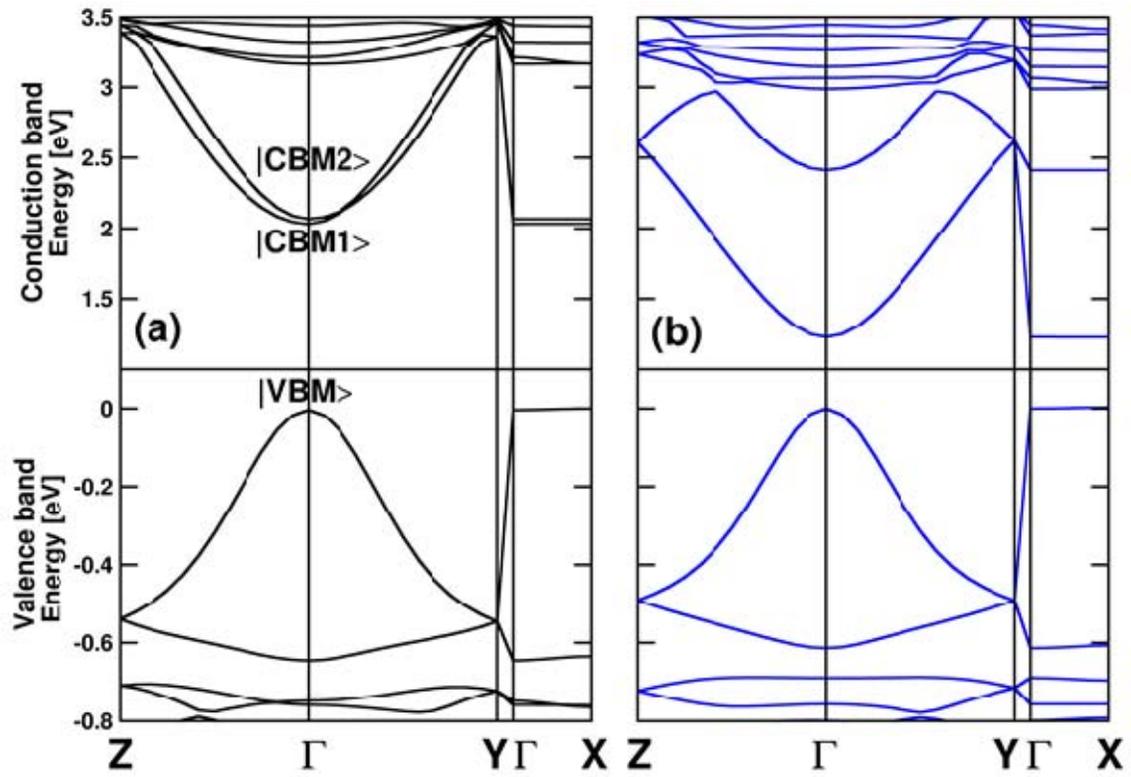

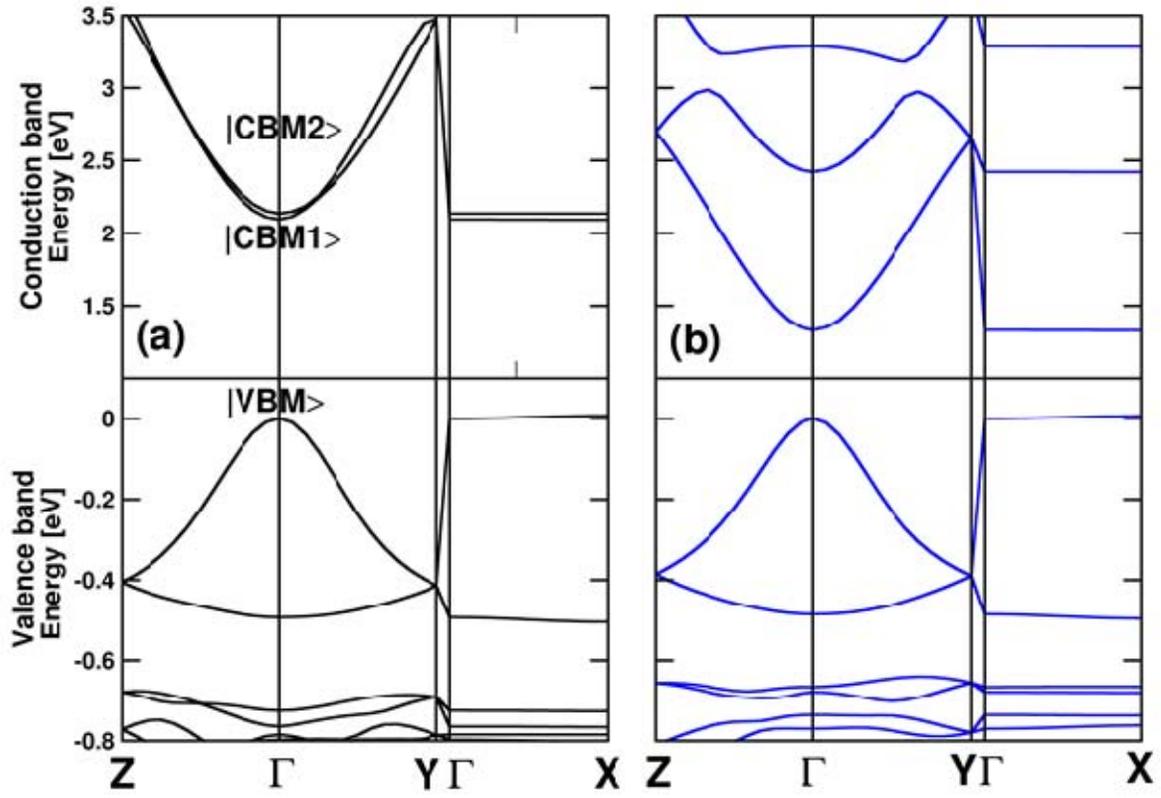

**FIGURE 4:** (J. Even for PRB)

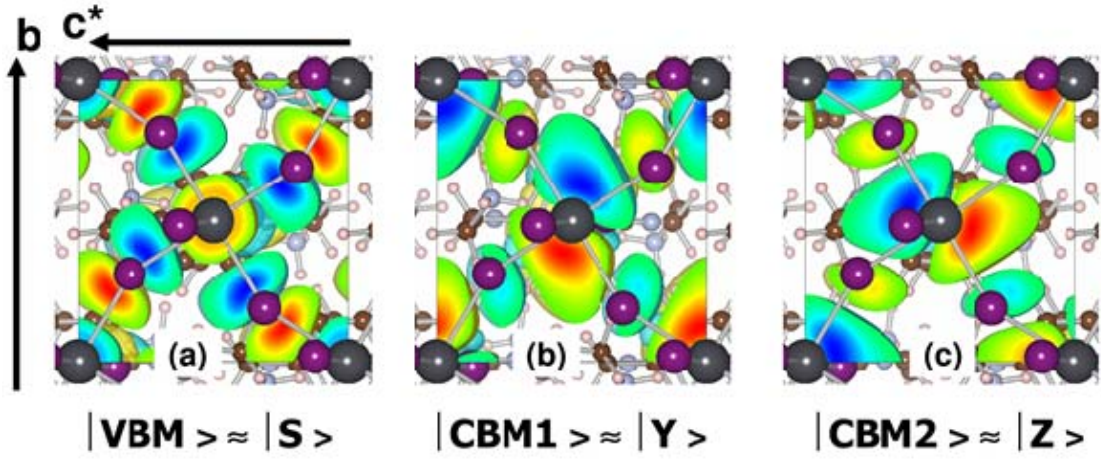

**FIGURE 5:** (J. Even for PRB)

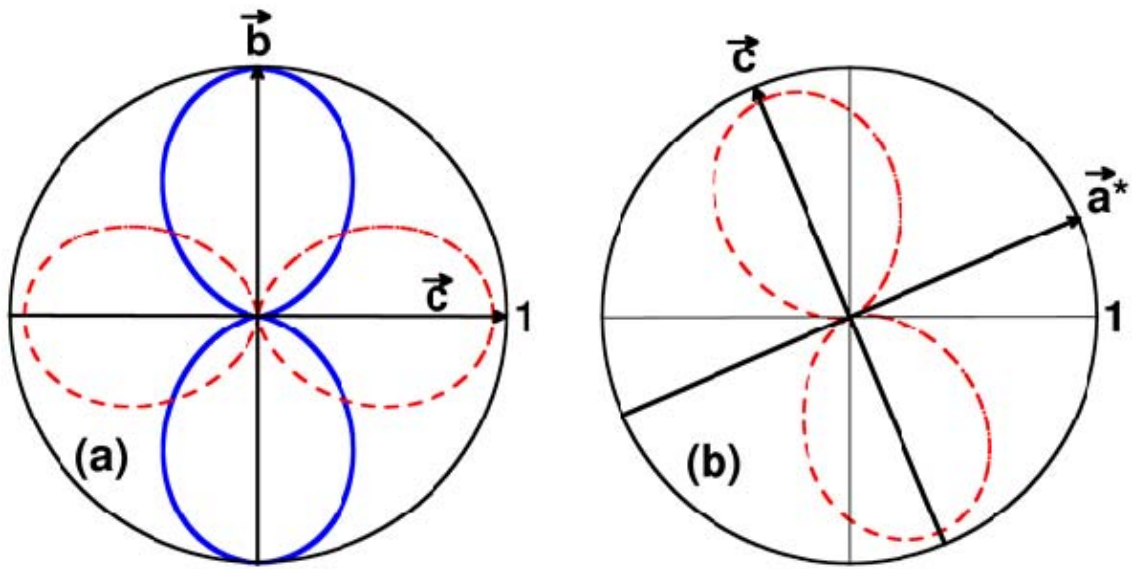

**FIGURE 6:** (J. Even for PRB)

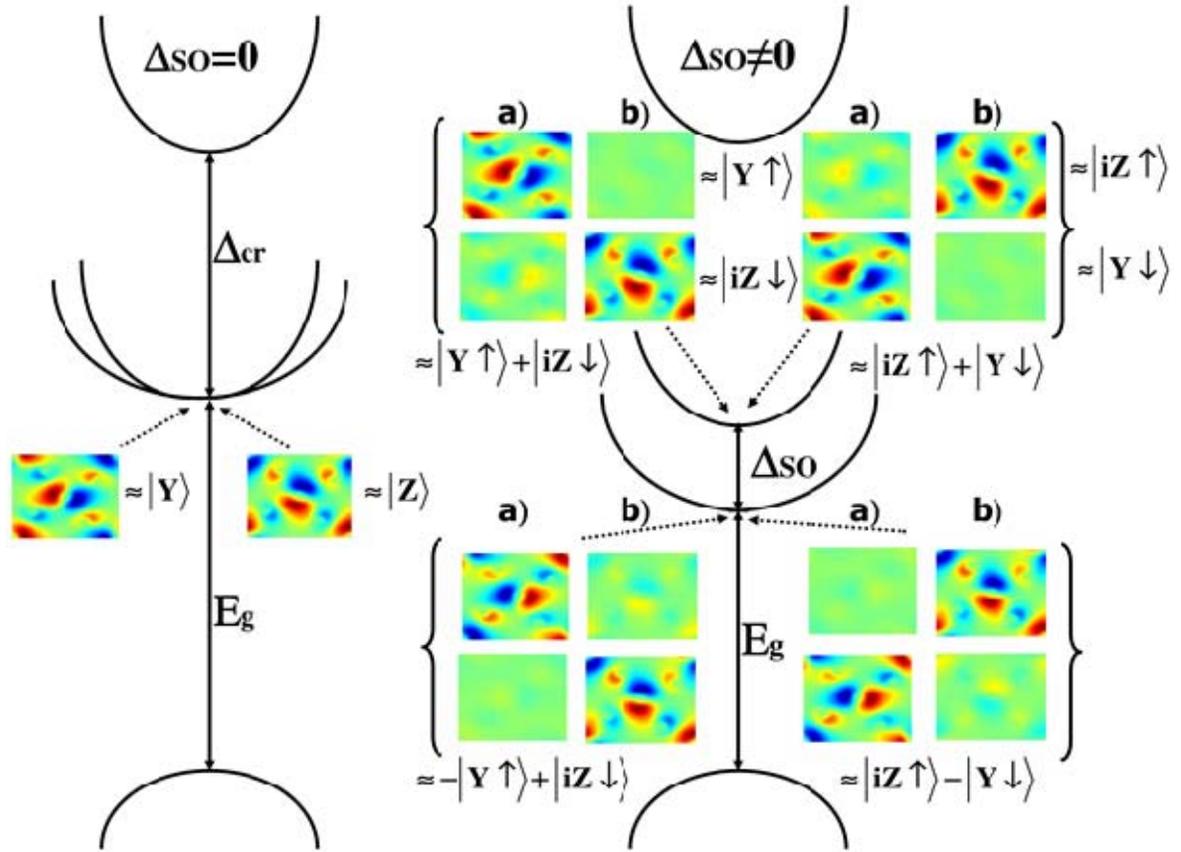

# FIGURE CAPTIONS

**Fig. 1:** (Color online) Overview of the crystal structure of 4F-PEPI ([pFC$_6$H$_5$C$_2$H$_4$NH$_3$]$_2$PbI$_4$).

**Fig. 2:** (Color online) Electronic band structure of 4F-PEPI without (a) and with (b) the spin-orbit coupling interaction calculated by the density functional theory. The energy levels are referenced to the valence band maximum.

**Fig. 3:** (Color online) Electronic band structure of [C$_5$H$_{11}$NH$_3$]$_2$PbI$_4$ without (a) and with (b) the spin-orbit coupling interaction calculated by the density functional theory. The energy levels are referenced to the valence band maximum.

**Fig. 4:** (Color online) Electronic wave-functions of the valence band maximum VBM and the first two excited states CBM1 and CBM2 as represented in the bc* -plane.

**Fig. 5:** (Color online) Dependence of the optical strengths $M^2_{VBM,CBJ}$ of 4F-PEPI on the direction of the light polarization for the VBM-CBM1 (solid line) and VBM-CBM2 (dotted lines) transitions. The value of $M^2_{VBM,CB1}$ along the b-axis is taken as reference both in the bc-plane (a) and in the plane perpendicular to the b axis (b).

**Fig. 6:** (Color online) Schematic representation of SAHOP electronic band diagram without ($\Delta_{SO}=0$, left) and with ($\Delta_{SO}\neq 0$, right) the spin-orbit coupling interaction $\Delta_{SO}$. $\Delta_{cr}$ represents the anisotropy of the crystal field. The real (a) and imaginary (b) parts of the complex spinorial components of the CBM1 and CBM2 states are represented for 4F-PEPI (the spin up component is on top of the spin down component).

# REFERENCES

[1] E. O. Kane, The k.p method, Chapter 3, Semiconductors and Semimetals Vol. 1, R. K. Willardson, A. C. Beer, ed., New York: Academic Press, (1966)

[2] S. Chuang, Physics of Optoelectronic Devices, J. W. Goodman ed., New York: Wiley, (1995)

[3] S. L. Chuang and C. S.Chang, Phys. Rev. B54, 2491 (1996)

[4] D.B.Mitzi, S. Wang, C.A. Field, C.A. Chess and A.M. Guloy, Science 267, 1473 (1995)

[5] J. Ishi, H. Kunugita, K. Ema, T. Ban, and T. Kondo, Appl. Phys. Lett. 77, 3487 (2000)

[6] A. Kojima, K. Teshima, Y. Shirai and T. Miyasaka, J. Am. Chem. Soc. 131, 6050 (2009)

[7] G. Lanty, A. Bréhier, R. Parashkov, J.S. Lauret and E. Deleporte, New J.Phys. 10, 065007 (2008)

[8] Y. Wei, J.-S. Lauret, L. Galmiche, P. Audebert and E. Deleporte, Optics Express 20, 10399 (2012)

[9] I. Koutselas, P. Bampoulis, E. Maratou, T. Evagelinou, G. Pagona and G.C. Papavassiliou, J. Phys. Chem. C 115, 8475 (2011)

[10] S. Zhang, G. Lanty, J. S. Lauret, E. Deleporte, P. Audebert and L. Galmiche, Acta Mater. 57, 3301 (2009)

[11] X. Hong, T. Ishihara and A. U. Nurmikko, Phys. Rev. B45, 6961 (1992)

[12] Y. Kato, D. Ichii, K. Ohashi, H. Kunugita, K. Ema, K. Tanaka, T. Takahashi and T. Kondo, Solid State Comm. 128, 15 (2003)

[13] M. Shimizu, J. Fujisawa and T. Ishihara, Phys. Rev. B74, 155206 (2006)

[14] K. Pradeesh, J. J. Baumberg and G. V. Prakash, Appl. Phys. Lett. 95, 173305 (2009)

[15] S. Sourisseau, N. Louvain, W. Bi, N. Mercier, D. Rondeau, F. Boucher, J.-Y. Buzaré and C. Legein, Chem. Mater. 19, 600 (2007)

[16] T. Ishihara, J. Takahashi and T. Goto, Phys. Rev. B42, 11099 (1990)

[17] K. Kikuchi, Y. Takeoka, M. Rikukawa and K. Sanui, Current Appl. Phys. 4, 599 (2004)

[18] X. Gonze, B. Amadon, P. M. Anglade, J. M. Beuken, F. Bottin, P. Boulanger, F. Bruneval, D. Caliste, R. Caracas, M. Cote, T. Deutsch, L. Genovese, Ph. Ghosez, M. Giantomassi, S. Goedecker, D. R. Hamann, P. Hermet, F. Jollet, G. Jomard, S. Leroux, M. Mancini, S. Mazevet, M. J. T. Oliveira, G. Onida, Y. Pouillon, T. Rangel, G. M. Rignanese, D. Sangalli, R. Shaltaf, M. Torrent, M. J. Verstraete, G. Zerah and J. W. Zwanziger, Computer Phys. Commun. 180, 2582 (2009)

[19] J. P. Perdew, K. Burke and M. Ernzerhof, Phys. Rev. Lett. 77, 3865 (1996)

[20] C. Hartwigsen, S. Goedecker and J. Hutter, Phys. Rev. B58, 3641 (1998)

[21] A. Lemmerer and D. G. Billing, Dalton Trans., 41, 1146 (2012)

[22] D. G. Billing and A. Lemmerer, Acta Cryst. B63, 735 (2007)

[23] S. L. Altmann and P. Herzig, Point-Group Theory Tables; Clarendon Press, Oxford (1994)

[24] T. Kataoka, T. Kondo, R. Ito, S. Sasaki, K. Uchida and N. Miura, Phys. Rev. B47, 2010 (1993)